\documentclass[10pt]{article}
\usepackage[OE]{express}

\begin{document}

\title{Suppression of Dick Effect in the Ramsey-CPT atomic clock by Interleaving Lock}

\author{X.L. Sun$^{1,2}$, J.W. Zhang$^{2,3,4}$, P.F. Cheng$^{1,2}$, Y.N. Zuo$^{1,2}$,  and L.J. Wang$^{1,2,3,5}$}

\address{
$^1$State Key Laboratory of Precision Measurement Technology and Instruments, Tsinghua University, Beijing 100084, China\\
$^2$Department of Physics, Tsinghua University, Beijing 100084, China\\
$^3$Department of Precision Instruments, Tsinghua University, Beijing 100084, China\\
$^4$zhangjw@tsinghua.edu.cn\\
$^5$lwan@tsinghua.edu.cn}

\email{} 



\begin{abstract}
For most passive atomic clocks, the Dick effect is one of the main limits to reach its frequency stability limitation due to quantum projection noise. In this paper, we demonstrate that the minimization of the Dick effect for the Ramsey-CPT atomic clock can be accomplished by interleaving lock. By optimizing the duty circle of laser pulses, averaging time during detection and optical intensity of laser beam, the Dick effect induced Allan deviation can be reduced to the level of 10$^{-14}$. 
\end{abstract}



\section{Introduction}
In recent years, the passive vapor cell frequency standards have made much progress and reached the frequency stability of 10$^{-13}\tau^{-1/2}$ or even better \cite{godone_physics_2006,micalizio_metrological_2012,lin_detection_2012,danet_compact_2014,noauthor_coherent_2015,dong_recent_2017,li_frequency_2014,peter_coherent_2016}. This kind of atomic clocks are suitable as secondary frequency standards where frequency stability, size and power consumption are of equal importance \cite{noauthor_coherent_2015}. Among those vapor cell standards, atomic clocks based on coherent population trapping (CPT) seem to be rather promising , due to the simple all optical setup and good performance \cite{liu_ramsey_2013}. The small signal amplitude of CPT clocks can be enhanced by optimized CPT pumping scheme \cite{taichenachev_high-contrast_2004,zanon_high_2005,jau_push-pull_2004,shah_high-contrast_2007}, and the line-width of clock transition can be reduced by Ramsey's method of separated oscillation fields \cite{zanon_high_2005}. However, the Dick effect is one of the main limits for the improvement of the frequency stability to the level of 10$^{-14}\tau^{-1/2}$. In 1987, Dick pointed out that the down-conversion of the local oscillator's frequency noise degrades the atomic clocks' frequency stability \cite{dick1987local}, and lots of work have been carried out to analyze and minimize Dick effect in the microwave atomic clocks \cite{dick1990local,audoin1998properties,santarelli1998frequency,greenhall1998derivation,biedermann_zero-dead-time_2013,li-jun_new_2014,jian-wei_dick_2015,chen_evaluation_2016} and optical atomic clocks \cite{quessada_dick_2003,westergaard_minimizing_2010,schioppo_ultrastable_2017}, respectively. Typically, there are two ways to reduce the Dick effect. One way is using an ultra-low phase noise local oscillator (LO), such as a cryogenic sapphire oscillator \cite{santarelli1999quantum} or an ultra-stable laser stabilized by an optical cavity\cite{PhysRevA.79.031803}. However, these oscillators are expensive and bulky. The other way is to interleaving lock the LO to two quantum systems \cite{dick1987local,biedermann_zero-dead-time_2013,jian-wei_dick_2015,schioppo_ultrastable_2017}. 

The analysis of Dick effect in the Ramsey-CPT atomic clocks is rather different from other atomic clocks, and methods to minimize it are seldom discussed. For the Ramsey-CPT clock, there are no straightforward  dead time and interrogation time as in the fountain clock, ion trap clock or optical lattice clock. The detection of clock signal takes place at the leading edge of one laser pulse, and atoms are prepared to the CPT dark states in the remnant part of the same laser pulse. Besides, the theoretical calculation of sensitivity function of Ramsey-CPT is based on a three-level atomic model rather than a simple two-level model. These make it more challenging to analyze and optimize the Ramsey-CPT clock to reduce Dick effect than other atomic clocks. 

In this paper, we present a detailed analysis of Dick effect in Ramsey-CPT clocks with interleaving lock by two vapor cells. We find that the sensitivity function for interleaved Ramsey-CPT clock is dependent on three parameters, the duty cycle of laser pulses, average time during detection and optical intensity of laser beam. According to our calculation, the Dick effect can be reduced from 6.11$\times$10$^{-13}\tau^{-1/2}$ to the level of 1.64$\times$10$^{-14}\tau^{-1/2}$ for the same LO.

\section{Dick effect in Ramsey-CPT atomic clock}
The Ramsey-CPT clock works in a pulse mode \cite{guerandel_raman_2007}. The time sequence of this clock is presented in Fig. 1(a). First, the dichromatic light lasting for a time $\tau_{L}$ pumps the atoms to the dark states. Then the light is absent to let the atoms freely process for a duration of T. Next, the atoms' states are detected by a following optical pulse. The transmitted light signal is measured at the end of $\tau_{m}$ and averaged for a time $\tau_{A}$. Atoms are then pumped to the dark states again. The cycle time is $T_{c} = \tau_{L}  + T$ in total.

\begin{figure}[ht!]
\centering\includegraphics[width=\textwidth]{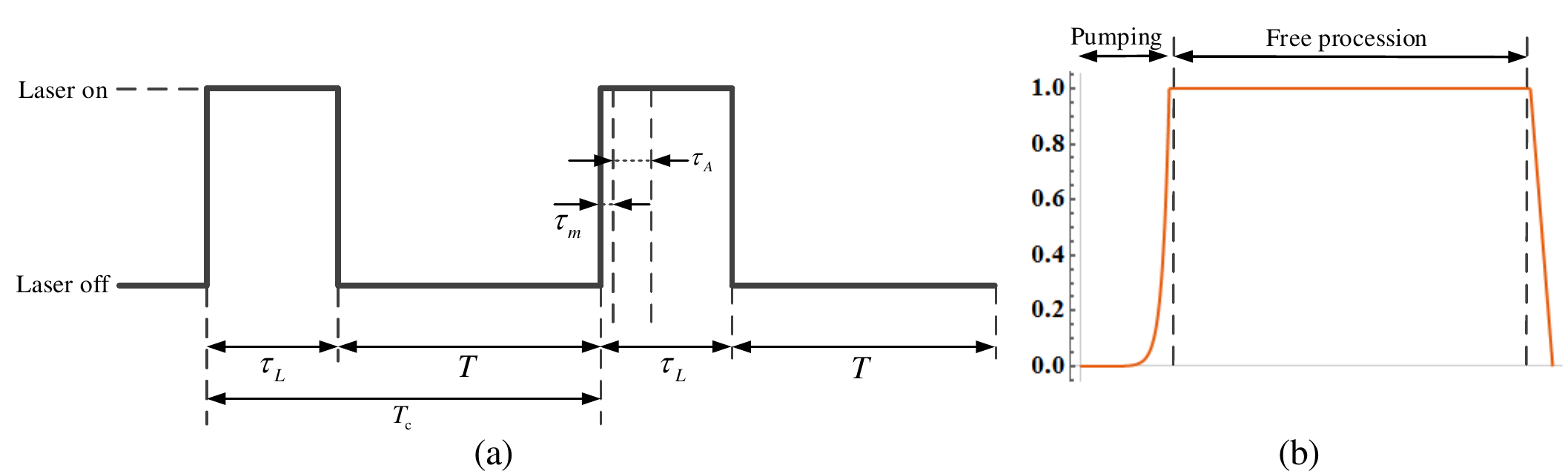}
\caption{(a) Time sequence of laser  pulses for a typical Ramsey-CPT atomic clock (b) Calculated sensitivity function $g(t)$ of a Ramsey-CPT system}
\end{figure}

The Dick effect induced Allan deviation (DEAD) of the locked LO can be expressed as \cite{dick1987local}
\begin{equation}
\sigma _y^{Dick}(\tau ) = {[\frac{1}{\tau }\sum\limits_{m = 1}^\infty  {(\frac{{g_{ms}^2 + g_{mc}^2}}{{g_0^2}})} S_y^f(\frac{m}{{{T_c}}})]^{1/2}},
\end{equation}
where $\tau$ is the averaging time, $S_y^f({m \mathord{\left/
 {\vphantom {m {{T_c}}}} \right.
 \kern-\nulldelimiterspace} {{T_c}}})$ is the one-side power spectral density (PSD) of the relative frequency fluctuations of the free-running LO at Fourier frequency ${m \mathord{\left/
 {\vphantom {m {{T_c}}}} \right.
 \kern-\nulldelimiterspace} {{T_c}}}$. $g_{ms}$, $g_{mc}$ and $g_{0}$ are respectively defined as 
\begin{equation}
\left( \begin{array}{l}
{g_{ms}}\\
{g_{mc}}
\end{array} \right) = \frac{1}{{{T_c}}}\int_0^{{T_c}} {g(\theta )} \left( \begin{array}{l}
\sin (2\pi m\theta /{T_c})\\
\cos (2\pi m\theta /{T_c})
\end{array} \right)d\theta 
\end{equation}
and
\begin{equation}
{g_0} = \frac{1}{{{T_c}}}\int_0^{{T_c}} {g(\theta )} d\theta,
\end{equation}
where $g(\theta )$ is the sensitivity function.

In the case of Ramsey-CPT, $g(t)$ can be expressed as\cite{danet_dick_2014}

\begin{equation}
g(t) = \left\{ \begin{array}{ll}
\exp {(\frac{{t - \tau_{L} }}{{{\tau _p}}})^{}}&(0 \le t \le \tau_{L} )\\
{1^{}}&(\tau_{L}  < t < \tau_{L}  + T + {\tau _m})\\
1 - {\frac{{t - (\tau_{L}  + T + {\tau _m})}}{{{\tau _A}}}^{}}&(\tau_{L}  + T + {\tau _m} \le t \le {T_c} + {\tau _m} + {\tau _A})
\end{array} \right.
\end{equation}
where ${\tau _p}{\rm{ = }}\Gamma /{\Omega ^2}$ is the pumping time, $\Gamma$ is the decay rate from the excited state, and ${\Omega ^2}$ is the quadratic sum of two optical transitions' Rabi frequencies. In our typical experiment condition, the $\tau{_p}$ is usually several microseconds. $g(t)$ is shown in Fig. 1 (b).  From the above equation, the pumping time is equivalent to the dead time as in other atomic clocks, during which frequency fluctuation of LO is not detected by the atoms.
\begin{figure}[ht!]
\centering\includegraphics[width=\textwidth]{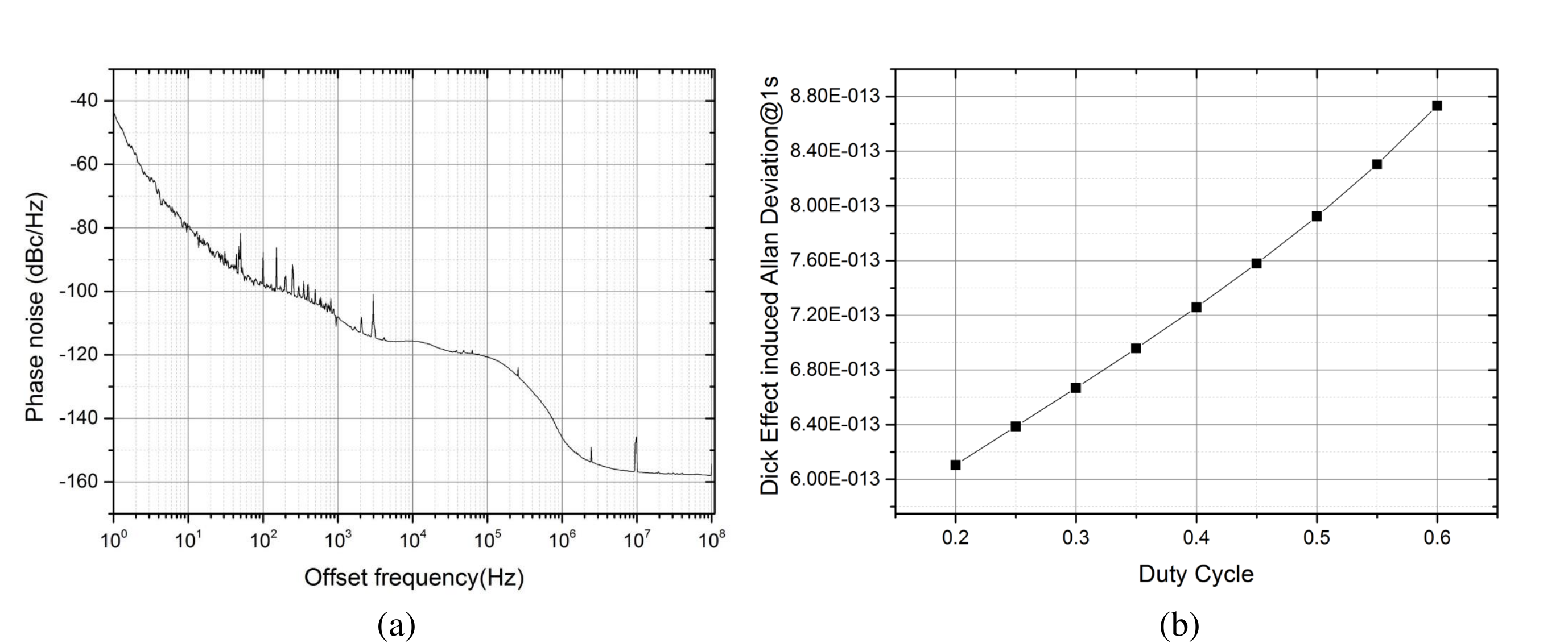}
\caption{(a) The measured phase noise of the 6.834GHz microwave synthesizer (b) Calculated Dick effect induced Allan deviation with 1s sampling time versus the duty cycle}
\end{figure}

In order to calculate the DEAD in our atomic clock, the PSD of the frequency fluctuations of the LO needs to be measured. In our experiment setup, the LO is a 10MHz oven controlled crystal oscillator (OCXO), and the microwave synthesizer is phase referenced to the LO to generate 6.834GHz microwave. The single-side phase noise of 6.8GHz is measured by a commercial signal source analyzer as shown in in Fig. 2 (a), and the PSD of the relative frequency can be expressed as
\begin{equation}
\begin{array}{l}
S_y^f(f) = 1.28 \times {10^{ - 24}}{f^{ - 2}} + 2.14 \times {10^{ - 26}}{f^{ - 1}} + \\
1.50 \times {10^{ - 26}}{f^0} + 6.42 \times {10^{ - 28}}{f^1} + 3.21 \times {10^{ - 32}}{f^2}.
\end{array}
\end{equation}
In the case of our CPT clocks, the free procession time is usually set to 800$\mu$s. Based on Eq. (1), the DEAD $\sigma _y^{Dick}(\tau )$ is calculated for different duty cycles while keeping the free evolution time unchanged. The calculation results are showed in Fig. 2 (b). As we can see, $\sigma _y^{Dick}(\tau )$ is 6.11$\times$10$^{-13}\tau^{-1/2}$ for our typical time sequence, and increases rapidly as the duty cycle increases. This result is very close to the previously estimated short-term frequency stability \cite{sun_investigation_2016,cheng_vapor_2016}. Thus, the Dick effect is a big issue for this clock to improve the short-term frequency stability below 1$\times$10$^{-13}\tau^{-1/2}$.
\section{Optimization of sensitivity function for interleaved Ramsey-CPT atomic clocks}
According to Eq. (1) and (2),  $g_{ms}$ and $g_{mc}$ are respectively the sine component and cosine component of sensitivity function $g(t)$ at Fourier frequency $m/T_{c}$, and determine the frequency weight of LO's PSD in the calculation of DEAD. If $g(t)$ is a constant function, $g_{ms}$ and $g_{mc}$ are equal to zero. In such an ideal situation, the Dick effect is completely eliminated. In the real operation of a single cell clock, $g(t)$ is zero over a considerable portion of one cycle due to the dead time, manifesting itself as a periodic function of $T_{c}$ and resulting in high value of $g_{ms}$ and $g_{mc}$. The interleaved operation of two quantum systems compensates the zero part of the sensitivity function and reduces $g_{ms}$ and $g_{mc}$ substantially, as shown in Fig. 3.
\begin{figure}[ht!]
\centering\includegraphics[width=\textwidth]{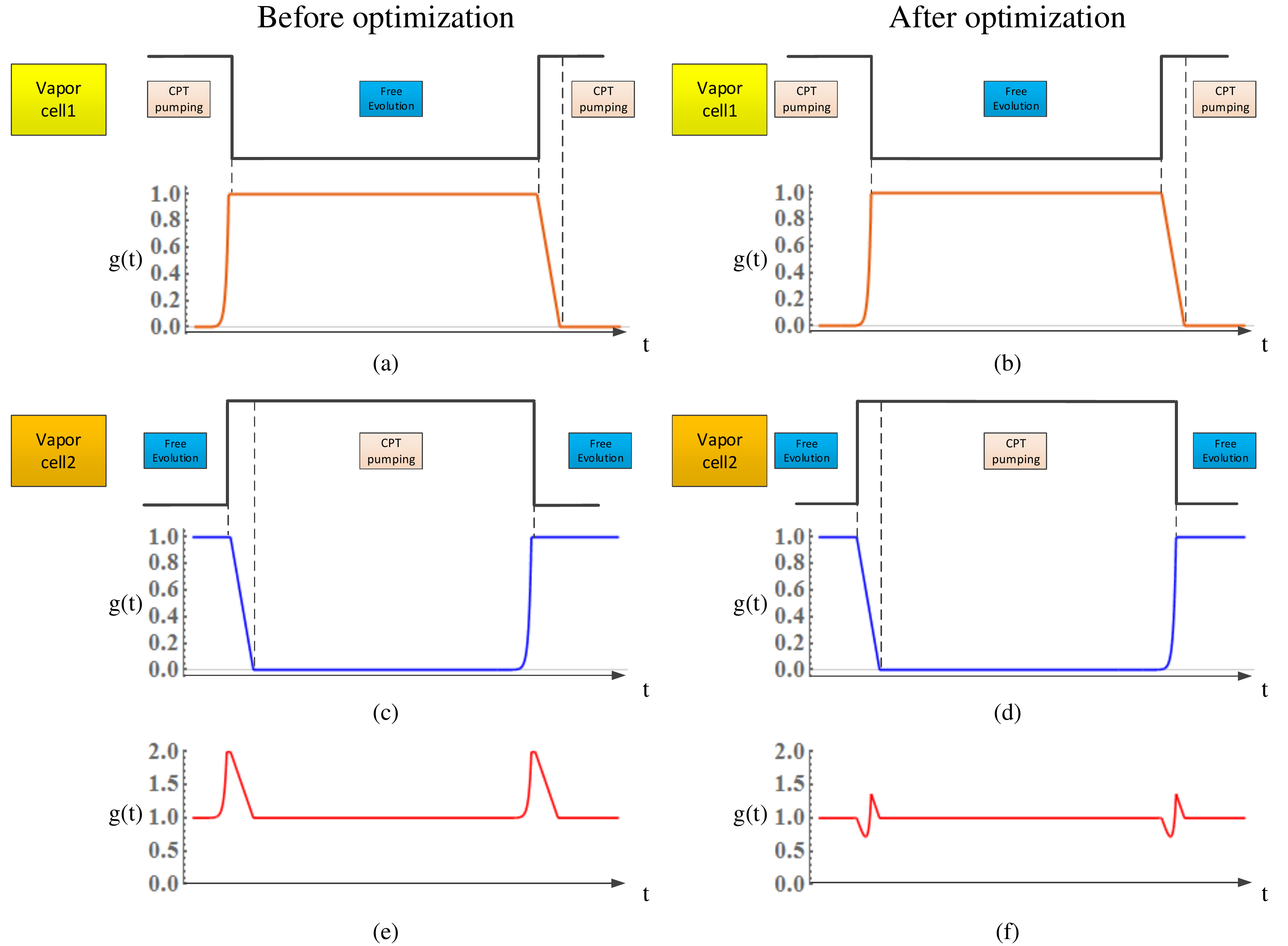}
\caption{Optimization of sensitivity function $g(t)$ for interleaved Ramsey-CPT clock.}
\end{figure}

Fig. 3 (a) and (c) show the time sequence and corresponding sensitivity function for vapor cell 1 and vapor cell 2, respectively. The duty cycle of the two pulsed laser beams are 50\%, and phase difference is 180\textdegree. Fig. 3 (e) shows the sum of the two sensitivity functions for the double cells. As we can see, although the summed sensitivity function is equal to unity in most of the cycle time, there still exist obvious high spikes at the beginning and ending of each cycle. Due to these periodic spikes, the sensitivity function is still a periodic function of $T_{c}$ and leads to non-zero $g_{ms}$ and $g_{mc}$. The calculated result of DEAD for the time sequence as shown in Fig. 3 (a) and (c) is 4.22$\times$10$^{-13}\tau^{-1/2}$, which is improved slightly compared with a single cell's value of 7.92$\times$10$^{-13}\tau^{-1/2}$. 

In order to obtain a more constant sensitivity function and reduce $g_{ms}$ and $g_{mc}$, the sequence can be optimized as shown in Fig. 3 (b) and Fig. 3 (d). The exponential rising part of the sensitivity function for one cell should partially overlap the linear falling for the other. When the areas above and below the line of $g(t)=1$ equals to each other, the time sequence is optimized to minimize the DEAD, as shown in Fig. 3 (f). As a result, the calculated DEAD decreases to 5.32$\times$10$^{-14}\tau^{-1/2}$, which is much improved compared to a single cell clock.

Besides duty cycle, averaging time during detection and optical intensity of laser beam are also of vital importance for the optimization of sensitivity function. The optical intensity is proportional to the CPT pumping rate and determines the rising edge of the sensitivity function. The averaging time determines the falling edge. First, we maintain the optical intensity constant and calculate the DEAD with different averaging times and duty cycles. The results are shown in Fig. 4 (a). As we can see, for different duty cycles, there always exists an average time to minimize the DEAD. By the calculation, the optimal duty cycle is 50.9\%, and the average time is 6$\mu$s for our setup. Based on these two values, the DEAD are calculated by varying different optical intensity, as shown in Fig. 4b. The optimal optical intensity is 9mW/cm$^{2}$, and the DEAD can be suppressed to 1.64$\times$10$^{-14}\tau^{-1/2}$, which is much less than the predicted Allan deviation limited by quantum projection noise of our atomic clock [28].

\section{Conclusion}
In summary, the sensitivity function of the Ramsey-CPT system is presented and the dead time during CPT pumping leads to degradation of clock's frequency stability. Based on the experimentally measured phase noise of the frequency synthesizer for 6.834GHz, we have successfully optimized the sensitivity function of the interleaved operating Ramsey-CPT atomic clock by adjusting duty cycle of laser pulses, average time during detection and optical intensity  of laser beam. Given the appropriate parameters, the Dick effect has been reduced to 1.64$\times$10$^{-13}\tau^{-1/2}$ for our setup. It can be hoped that interleaved operating Ramsey-CPT atomic clock will be a high-performance and compact atomic clock.
\begin{figure}[ht!]
\centering\includegraphics[width=\textwidth]{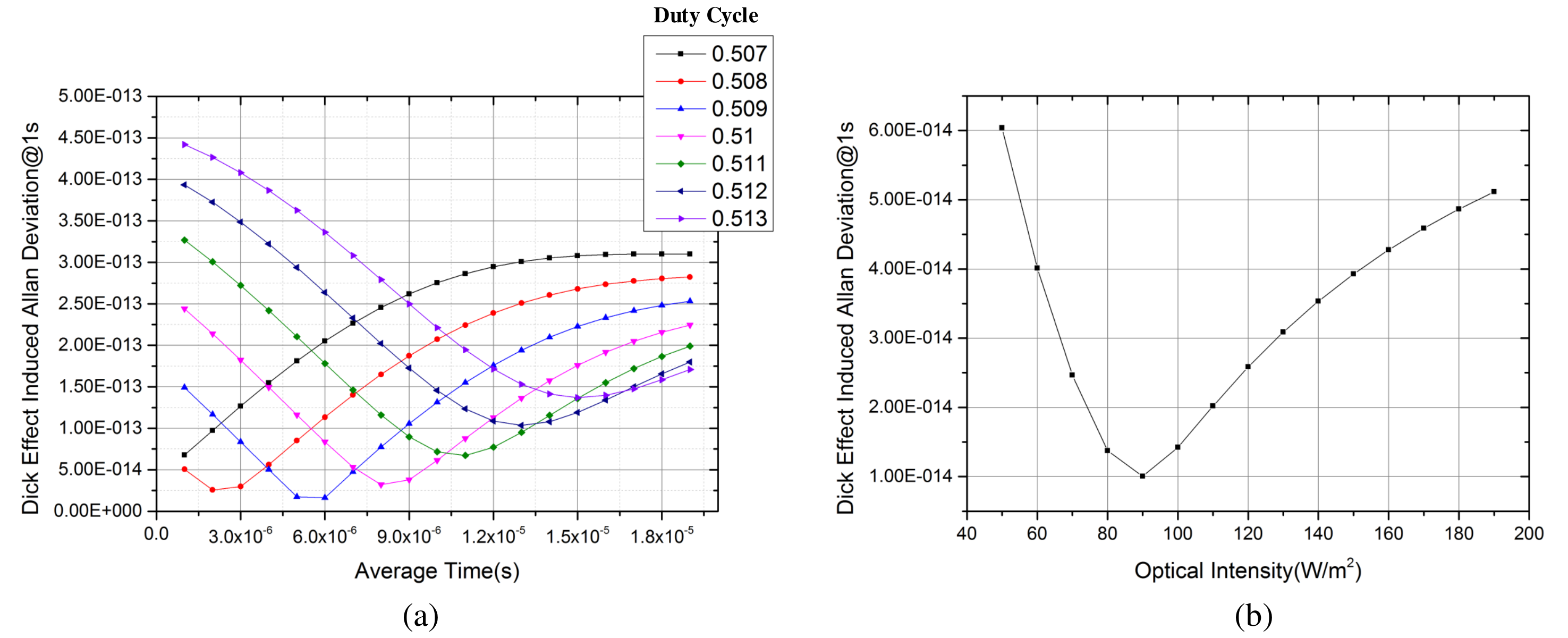}
\caption{(a) Optimization of sensitivity function versus duty cycle and average time (b) Optimization of sensitivity function versus optical intensity (duty cycle is 50.9\% with an average time of 6$\mu$s)}
\end{figure}
\section*{Funding}
This work is supported by the National Key Research and Development Program of China (No. 2016YFA0302101), the National Natural Science Foundation of China (No. 11304177), and the Initiative program of State Key Laboratory of Precision Measurement Technology and Instruments.
\section*{Acknowledgments}
We would like to thank Chi Xu and Chenfei Wu for helpful discussions. 
\end{document}